\documentclass[aps,prc,superscriptaddress,twoside,twocolumn,nofootinbib,showpacs]{revtex4}
\usepackage{amsmath,amssymb}
\usepackage{mathtools}
\usepackage{bbold}
\usepackage[usenames, dvipsnames]{xcolor}
\usepackage{braket}
\usepackage{graphicx,bm}
\usepackage{slashed}
\usepackage{booktabs}
\usepackage{tensor}
\usepackage[utf8]{inputenc}

\usepackage{multirow}

\usepackage{changes}

\usepackage{soul}

\usepackage{enumerate}

\definechangesauthor[name={Yeunhwan}, color=blue]{YH}

\allowdisplaybreaks

\usepackage{pgfplots}
\usepackage{xparse}

\begin{document}

\title{\boldmath Proton pairing in neutron stars from chiral effective field theory}
\date{\today}

\author{Yeunhwan \surname{Lim} }
\affiliation{Cyclotron Institute, Texas A\&M University, College Station, TX 77843, USA}

\author{Jeremy W. \surname{Holt} }
\affiliation{Cyclotron Institute, Texas A\&M University, College Station, TX 77843, USA}
\affiliation{Department of Physics and Astronomy, Texas A\&M University, College Station, TX 77843, USA}

\begin{abstract}
We study the ${}^{1}S_0$ proton pairing gap in beta-equilibrated neutron star matter within
the framework of chiral effective field theory. We focus on the role of three-body forces, which
strongly modify the effective proton-proton spin-singlet interaction in dense matter. We find that
three-body forces generically reduce both the size of the pairing gap and the maximum density 
at which proton pairing may occur. The pairing gap is computed within BCS theory using a
single-particle dispersion relation calculated up to second order in perturbation theory. Model
uncertainties are estimated by varying the nuclear potential (its order in the chiral expansion
and high-momentum cutoff) and the choice of single-particle
spectrum in the gap equation. We find that a second-order perturbative treatment of the 
single-particle spectrum suppresses the proton ${}^{1}S_0$ pairing gap relative to the 
use of a free spectrum.
We estimate the critical temperature for the onset of proton superconductivity to be 
$T_c = (3.2 - 5.1)\times 10^{9} $\,K, which is consistent with previous theoretical results in the 
literature and marginally within the range deduced from a recent Bayesian analysis of neutron star 
cooling observations.
\end{abstract}


\maketitle

\section{Introduction}\label{sec:int}

Neutron superfluidity and proton superconductivity play an important role in the physics of
neutron stars \cite{dean03}. 
The dilute gas of neutrons in the inner crust of a neutron star is expected to
pair in the spin-singlet channel, resulting in a neutron superfluid whose vortices provide a large
angular momentum reservoir critical for the production of pulsar glitches 
\cite{BAYM1969,andersson12,chamel12,piekarewicz14}. 
At the higher densities present in the core of neutron stars, the proton fraction is much less 
than that of neutrons, and therefore the proton Fermi momentum is not comparable with 
the neutron Fermi momentum and the formation of neutron-proton Cooper pairs is unlikely. 
It is then natural to consider neutron-neutron and proton-proton pairing
separately. At large densities, neutrons may be paired 
in the spin-triplet channel, leading to novel cooling processes involving pair formation/breaking
that can impact the early thermal evolution of neutron stars 
\cite{FRS1976,VS1987,Yakovlev:1999sk,Page:2004fy,Heinke10,Page:2009fu,Page:2010aw,Shternin:2010qi,Lim:2015lia}. 
Neutron star cooling may also be affected by the presence of superconducting protons in 
neutron star cores \cite{Kaminker99,Gusakov04}, though the critical temperature is expected 
to be somewhat larger than that
for neutron superfluidity and consequently would impact the cooling curve at earlier 
timescales. Well below the critical temperature for neutron superfluidity and proton superconductivity, 
neutrino emission involving neutrons or protons is highly suppressed due to the minimum gap
energy required to break a Cooper pair\,\cite{yakovlev01}. 
Superfluidity also gives reduction factors to the heat capacity and thermal conductivity
of dense nuclear matter \cite{Yakovlev:1999sk,baiko01}.
Besides neutrino emission from Cooper pairs, 
superfluidity and superconductivity in the crust and core affect pulsar glitches 
\cite{Link1992,ho2015}, vortex pinning \cite{alpar84,epstein88},
and neutron star precession \cite{Link03,Glampedakis08}.

Accurate estimates for nuclear pairing 
gaps in the various spin and isospin channels are challenging due to uncertainties in strong 
interaction physics, in particular poorly constrained nuclear many-body correlations and 
three-body forces that become increasingly important at high densities.
In the past, neutron spin-singlet pairing in pure neutron matter
has been widely studied, with recent work focusing on
the role of three-body forces \cite{Hebeler10,Maurizio14} and long- and short-range correlations 
\cite{Cao06,Ding16} in the BCS approximation. Quantum Monte Carlo calculations \cite{Gezerlis08},
on the other hand, can explore neutron pairing in the strong superfluid regime and
connections to ultracold Fermi gases at unitarity.
In nearly all cases, however, lattice effects and the presence of nuclear clusters in the neutron
star crust are neglected in microscopic many-body calculations of the neutron $^1S_0$ pairing gap. 
Spin-triplet pairing of neutrons in the neutron star core is anticipated from the strong attraction
in the ${^3P_2}-{^3F_2}$ partial-wave channel observed in nucleon-nucleon (NN) elastic scattering~\cite{Stoks93}. However, many-body effects 
such as screening, short-range correlations, and three-body forces play a substantial role, 
and there is currently much uncertainty in estimates of the spin-triplet pairing gap 
(for a recent review, see Ref.\ \cite{Gezerlis14}).

Previous works 
\cite{CHAO1972320,Tak73,Amundsen:1984qq,BALDO1992349,CHEN199359,Elgaroy:1996mx} 
studying proton pairing in neutron star cores have employed a variety of NN interaction models 
and many-body methods. The peak in the proton pairing gap was found to vary between 
$\Delta \simeq 0.4-0.9$\,MeV and to occur around
normal nuclear densities $n_0 \simeq 0.16$\,fm$^{-3}$, though the density of protons is one or two orders of 
magnitude less and set by the condition of beta equilibrium. More recently \cite{ZUO200444} a three-body 
force based on $\pi$ and $\rho$ meson exchange was included in the solution of the BCS gap equation 
and found to reduce by half the maximum value of the proton pairing gap compared to the inclusion of 
two-body forces alone. The two-body force employed in Ref.\ \cite{ZUO200444} was the Argonne 
$v_{18}$ potential, which includes explicit one-pion exchange at large distances but treats the 
medium- and short-range parts of the NN potential in terms of parametrized 
phenomenological functions.

In the present study we focus on a microscopic description of proton 
pairing in neutron star cores employing a set of two- and three-body nuclear forces 
\cite{entem03,marji13,coraggio13,coraggio14,sammarruca15} derived in 
the framework of chiral effective field theory \cite{weinberg79,epelbaum09rmp,machleidt11}.
Previous works employing these potentials have shown that they provide a good description of 
nuclear matter saturation \cite{coraggio14,sammarruca15}, the liquid-gas phase transition 
\cite{wellenhofer14,holt13ppnp,holt16pr}, nucleon-nucleus optical 
potentials \cite{holt13prc,holt16prc} and Fermi liquid parameters \cite{holt12npa}. In addition
the derived nuclear
equation of state (EOS) is consistent with other studies 
\cite{bogner05,hebeler11,hagen14,carbone14,drischler16}
employing different chiral nuclear forces and many-body methods. The present work will be
important for developing consistent modeling of the equation of state and nucleonic 
pairing needed for neutron star cooling calculations.

The paper is organized as follows. In Section \ref{ge} we describe the method
employed to solve the BCS gap equation. We also detail the treatment of the proton-proton
effective interaction and the proton single-particle potential in neutron-rich matter from chiral effective field
theory. In Section \ref{results} we present results for the density-dependent $^1S_0$ proton pairing 
gap at the Fermi surface in beta-equilibrated nuclear matter. Theoretical uncertainties are estimated 
by varying the 
resolution scale of the nuclear potential, the order in the chiral expansion, and the treatment of
the single-particle dispersion relation. We conclude with a summary and outlook.

\section{Proton pairing gaps in neutron-rich matter}
\label{ge}

\subsection{BCS gap equation}

The $^{1}S_0$ pairing gap for a given baryon number density 
can be obtained in the BCS approximation by solving the gap equation
\begin{equation}
\label{eq:gap}
\Delta(k) =  -\frac{1}{2}\sum_{k^\prime} V_{\rm eff}(k,k^\prime)
\frac{\Delta(k^\prime)}{\sqrt{(e_{k^\prime} - \mu)^2 + \Delta^{2}(k^\prime)}} \,,
\end{equation}
where $\Delta(k)$ is the pairing gap for the momentum $k$,
$V_{\rm eff}(k,k^\prime)$ is the effective potential between two incoming particles with 
relative momentum $k$ and outgoing relative momentum $k^\prime$. The single-particle energy as a 
function of momentum $k$ is denoted by $e_k$, and $\mu$
is the chemical potential for protons at a given density.
In the BCS approximation,
the effective potential is chosen as the free-space nucleon-nucleon interaction $V_{NN}$. Improved 
approximations account for medium effects, such as three-body force contributions to the in-medium
NN interaction $V_{NN}^{\rm med}$, as well as long-range correlations and polarization effects. In the present work,
we will consider only the additional contributions to $V_{\rm eff}(k,k^\prime)$ from three-body
forces. Extensions to include polarization effects through the Fermi liquid theory quasiparticle
interaction \cite{Ding16,sedrakian19,holt18} will be studied in later work.

Many BCS calculations in nuclear matter employ an effective mass approximation
\begin{equation}
\label{eq:effm}
e_k = \frac{k^2}{2M^*}+U,
\end{equation}
where $U$ depends on the density but is independent of the momentum $k$. 
From Eq.\ \eqref{eq:effm}, the gap equation is then approximated by substituting
\begin{equation}
e_k - \mu \simeq \frac{1}{2M^*}(k^2-k_F^2)\,.
\end{equation}
The above approximation assumes that the single-particle energy is nearly quadratic
in $k$ near the Fermi momentum $k_F$.
In this case the numerical solution for Eq.~\eqref{eq:gap} can be obtained from a generalized matrix 
eigenvalue solution~\cite{fan2015} by noting that $\Delta_i = F_{ij} \Delta_j $ as in gap 
Eq.\,\eqref{eq:gap}. In practice one applies an adaptive mesh point scheme that depends 
on the effective mass for a given Fermi momentum.

We also employ the modified Broyden method \cite{Johnson1988} to verify 
our numerical solutions. In this approach the gap solution is obtained from a version of 
direct iteration, where an initial guess of the momentum-dependent gap function is
inserted into the right-hand side of Eq.\ (\ref{eq:gap}) to obtain an updated guess for the 
gap function on the left-hand side of Eq.\ (\ref{eq:gap}). The modified Broyden method uses 
a numerically efficient algorithm for computing and storing the Jacobian, from
which the gap equation can be solved iteratively using a pseudo-Newton convergence method
(see Refs.\ \cite{Johnson1988, Drischler17} for complete details).
We find that this method converges rapidly once we have an initial guess for 
the gap size $\Delta(k)$. Moreover, the solution is not particularly sensitive to the 
initial guess $\Delta^{(0)}(k)$.
We find that both methods agree within $1$~keV 
when we use the effective mass approximation in Eq.\ (\ref{eq:effm}).

The numerical solution to the generalized matrix eigenvalue problem, however, is not
applicable when we use a general single-particle energy spectrum instead of the 
effective mass approximation. The matrix eigenvalue method enables us to obtain
an analytic solution for the denominator in Eq.\ \eqref{eq:gap}, 
$\sqrt{(e_{k^\prime} - \mu)^2 + \Delta^{2}(k^\prime)}$, when we apply the 
effective mass approximation. When the single particle energy spectrum does not 
actually follow a quadratic approximation, the different momentum-dependent effective 
masses give different gap sizes even though the order of magnitude is similar for each 
effective mass. Thus, we implement the Broyden method technique 
as explained in Drischeler \textit{et al}.\,\cite{Drischler17} to
obtain the BCS solution in this work. 
\begin{figure}[t]
\includegraphics[scale=0.5]{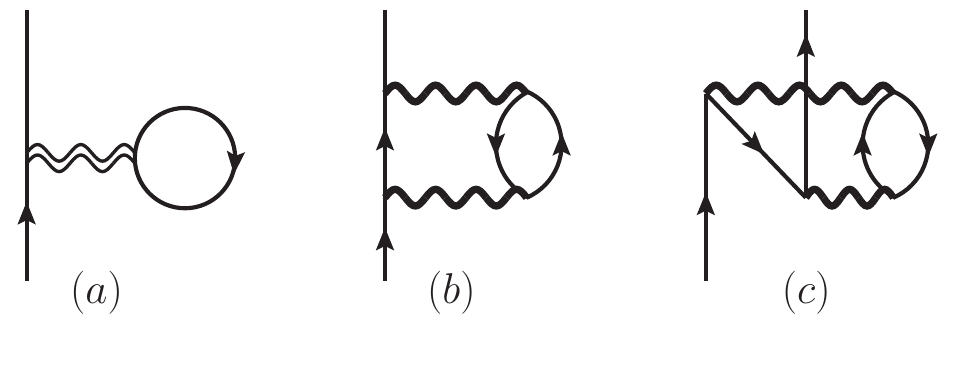}
\caption{Diagrammatic contributions to the nucleon self-energy in nuclear matter. 
The wavy line represents a medium-dependent effective NN
interaction derived from two- and three-body chiral forces in isospin-asymmetric
nuclear matter.}
\label{optpot}
\end{figure}

\subsection{Nucleon single-particle energy}

The single-particle energy spectrum plays an important role in determining the solution
to the gap equation, and
in the present work we consider three approximations to estimate the 
associated theoretical uncertainty. First, we assume a free-particle spectrum given by
$e^{(0)}_k = k^2/2M$. Second, we compute the proton single-particle energy
in the Hartree-Fock approximation
\begin{equation}
e^{(1)}_k = k^2/2M + \Sigma^{(1)}(k),
\label{sig1}
\end{equation}
where the first-order contribution $\Sigma^{(1)}(k)$ to the nucleon self energy is
shown diagrammatically in Fig.\ \ref{optpot}(a).
Third, we compute the single-particle energy self-consistently at second-order in 
perturbation theory
\begin{equation}
e^{(2)}_k = k^2/2M + \Sigma^{(1)}(k) + {\rm Re} \Sigma^{(2)}(e^{(2)}_k,k),
\label{se2}
\end{equation}
where $\Sigma^{(2)}(e_k,k)$ is represented by the sum of
diagrams (b) and (c) in Fig.\ \ref{optpot}. The first- and second-order
diagrammatic contributions to the nucleon self energy have the form

\begin{eqnarray}
&&\hspace{-.3in} \Sigma^{(1a)}_t(k) \\ \nonumber
&&\hspace{-.15in}= \sum_{1} \langle \vec k \, \vec h_1 s s_1 t t_1 | 
(\bar V_{NN}+ \bar V_{NN}^{\rm med}/2) | \vec k \,
\vec h_1 s s_1 t t_1 \rangle n_1,
\end{eqnarray}

\begin{eqnarray}
&&\hspace{-.3in}\Sigma^{(2b)}_t(k,\omega) \\  \nonumber
&&\hspace{-.15in}= \frac{1}{2}\sum_{123} \frac{| \langle \vec p_1 \vec p_3 s_1 s_3 t_1 
t_3 | \bar V_{\rm eff} | \vec k \, \vec h_2 s s_2 t t_2 \rangle |^2}{\omega + \epsilon_2 - \epsilon_1
-\epsilon_3 + i \eta} \bar n_1 n_2 \bar n_3,
\label{op2ac}
\end{eqnarray}

\begin{eqnarray}
&&\hspace{-.3in}\Sigma^{(2c)}_t(k,\omega) \\ \nonumber
&&\hspace{-.15in}= \frac{1}{2}\sum_{123} \frac{| \langle \vec h_1 \vec h_3 s_1 s_3 t_1 
t_3 | \bar V_{\rm eff} | \vec k \, \vec p_2 s s_2 t t_2 \rangle |^2}{\omega + \epsilon_2 - \epsilon_1
- \epsilon_3 - i \eta} n_1 \bar n_2 n_3,
\label{op2bd}
\end{eqnarray}
where $t$ labels the isospin quantum number of the external particle, 
$n_j=\theta(k_f-|\vec p_j|)$ is the zero-temperature momentum distribution function, 
$\bar n_j=1-n_j$, $\bar V= V-{\cal P}_{12}V$ is the antisymmetrized NN potential
with ${\cal P}_{12}$ the exchange-operator, and all sums are taken over momentum, 
spin, and isospin states.

The effective interaction in Eqs.\ \eqref{op2ac} and \eqref{op2bd} 
is defined by $V_{\text{eff}} = V_{NN} + V_{NN}^{\rm med}$, 
where $V_{NN}^{\rm med}$ is the density-dependent NN potential derived from the 
N2LO chiral three-body force by averaging one state over the 
filled Fermi sea of noninteracting protons and neutrons in asymmetric nuclear matter 
\cite{Sammarruca15b} (for additional details see Refs.\ \cite{Holt10,Hebeler10}).
In computing the in-medium interaction
$V_{NN}^{\rm med}$, we effectively normal order with respect to a noninteracting (unpaired)
ground state. The inclusion of pairing correlations in the normal-ordering reference
state for $V_{NN}^{\rm med}$ \cite{papa17} amounts to summing the third particle over a 
single-particle BCS spectrum but which otherwise leads to a gap equation that 
has the same structure as Eq.\ (\ref{eq:gap}). The double-wavy line in Fig.\ \ref{optpot}(a) 
represents the fact that in the first-order
Hartree-Fock contribution to the nucleon self-energy, there is an additional symmetry
factor of $\frac{1}{2}$ for the medium-dependent potential, namely $V_{\rm eff}^{HF} = V_{NN} 
+ \frac{1}{2}V_{NN}^{\rm med}$. The effective interaction
in the BCS gap equation, Eq.\ (\ref{eq:gap}), however, requires no additional 
symmetry factor \cite{Drischler17}.
We note that $V_{NN}^{\rm med}$ depends on both the density and composition,
namely, the proton fraction. The proton fraction is determined by enforcing 
beta equilibrium, which requires computing the proton and neutron chemical
potentials from the equation of state of asymmetric nuclear 
matter \cite{wellenhofer15,drischler16}. The electrons are treated as a relativistic gas of noninteracting
Fermions.

We employ chiral nucleon-nucleon interactions at next-to-next-to-leading order (N2LO) 
and next-to-next-to-next-to-leading order (N3LO) in the chiral power counting. 
For values of the momentum-space cutoff 
$\Lambda \lesssim 500$\,MeV, nucleon-nucleon potentials generally exhibit good convergence
properties in many-body perturbation theory. In the present work we therefore consider two
values of the cutoff ($\Lambda = 450$\,MeV and $500$\,MeV) at N2LO and three
values of the cutoff ($\Lambda = 414$\,MeV, $\Lambda = 450$\,MeV and $500$\,MeV) 
at N3LO \cite{marji13}. We note that the value $\Lambda = 414$\,MeV is not the result of 
fine tuning but instead corresponds to the relative momentum for nucleon-nucleon scattering 
at a lab energy of $E_{\rm lab} = 350$\,MeV, the maximum energy at which modern
nucleon-nucleon potentials are fitted to phase shifts. 
In all cases we include also the N2LO chiral three-body force whose
low-energy constants $c_D$ and $c_E$ are fitted to reproduce the binding energies 
of $A=3$ nuclei and the beta-decay lifetime of $^3$He \cite{coraggio14,sammarruca15}.
We note that in all cases we employ the charge-dependent versions of these potentials
that differ primarily in the leading-order low-energy constant associated with the ${}^1S_0$
partial wave.

The same approximations for the single-particle energy employed
in the present work have been shown to give a good description of the nucleon-nucleus
optical potential, especially the dependence of the real part on the isospin asymmetry and 
energy \cite{Whitehead:2018bfs,Whitehead:2019poy}. 
As we demonstrate below, the many-body 
perturbation series expansion of the single-particle energies appears to be under control,
but uncertainties persist. Despite the above consistencies in the treatment of the 
effective interaction and single-particle spectrum, additional many-body effects beyond
the BCS mean field approximation are important. In particular, both short- and long-range
correlations lead to a fragmentation of single-particle strength, encoded in nuclear 
spectral functions, that modify the quasiparticle energy spectrum. While such effects
are partly accounted for through our use of the single-particle energy at second
order in perturbation theory, Eq.\ (\ref{se2}), a complete treatment involving the
superfluid Green's function involves a more complicated double energy convolution
of the spectral function \cite{Rios:2017muz}. Short-range correlations have been
shown \cite{Ding16} to reduce by about 25\% the size of the neutron pairing gap 
in the spin-singlet channel. Long-range correlations in the effective pairing 
interaction, which represent the exchange of virtual collective modes, tend to 
decrease the strength of the singlet pairing gap by about 20\% or less on 
average for a range of nuclear interactions and densities \cite{Ding16}.

\section{Results}
\label{results}

\begin{figure}[t]
\includegraphics[scale=0.58]{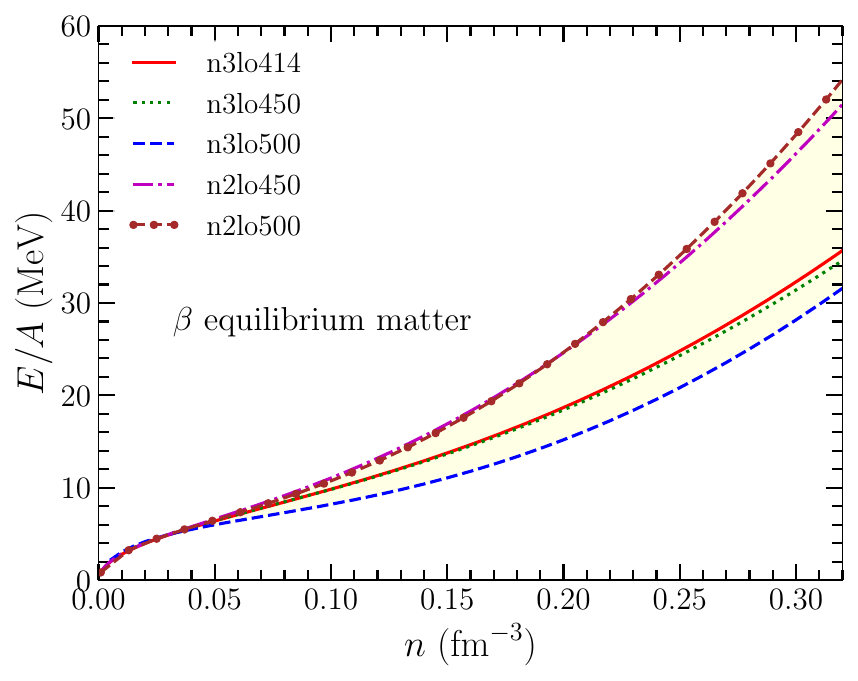}
\caption{(Color online) Equation of state of nuclear matter in beta equilibrium from 
the chiral two- and three-nuclear forces used in this work.}
\label{fig:eos}
\end{figure}

In Fig.\ \ref{fig:eos} we show the equation of state of beta equilibrated nuclear matter 
calculated from the five chiral nuclear forces employed in the present work. We first compute the 
equation of state for isospin-asymmetric nuclear matter at second order in perturbation
theory:

\begin{equation}
{\cal E}^{(1)} = \frac{1}{2}\sum_{12} \,
n_1 n_2 \langle 1 2 \vert (\bar V_{NN}+ \bar V_{NN}^{\rm med}/3)\vert 1 2 \rangle,
\label{e1}
\end{equation}

\begin{equation}
{\cal E}^{(2)} = -\frac{1}{4} 
\sum_{1234} \left| \langle 1 2 \left | \bar V_{\text{eff}} \right | 3 4 \rangle \right |^2
\frac{n_1 n_2 \bar{n}_3 \bar{n}_4}
{e_3+e_4-e_1-e_2},
\label{e2}
\end{equation}
\noindent where ${\cal E} = E/V$ is the energy density and the single-particle
energies $e_i$ in ${\cal E}^{(2)}$ are computed according to Eq.\ (\ref{sig1}).
Analogous to the calculation of the nucleon self energy in the previous section, 
the in-medium nucleon-nucleon interaction $V_{NN}^{\rm med}$
requires an additional symmetry factor of 1/3 in the calculation of the 
Hartree-Fock contribution to the energy density.

From Eqs.\ (\ref{e1}) and (\ref{e2}), the proton and neutron chemical potentials 
can be evaluated as
\begin{equation}
\mu_p = \left . \frac{\partial {\cal E}}{\partial n_p} \right |_{n_n}, 
\hspace{.3in} \mu_n = \left . \frac{\partial {\cal E}}{\partial n_n} \right |_{n_p},
\label{chemp}
\end{equation}
where $n_p$ is the proton number density and
$n_n$ is the neutron number density. The electron density is set by charge neutrality,
and beta equilibrium is then found by enforcing $\mu_n = \mu_p + \mu_e$. 
As a practical approach, we fit an energy density functional that 
is consistent with the chiral effective field theory 
neutron matter and symmetric nuclear matter equations of state
from many-body perturbation theory. We have verified that this introduces no significant
error in computing the chemical potentials. Strictly speaking, our perturbation theory 
treatment of the equation of state and single-particle potential does not constitute a 
conserving approximation, which means that there is some ambiguity in the definition
of the chemical potential. Nevertheless, we find good numerical agreement 
between the chemical potentials computed from Eq.\ (\ref{chemp}) and from the
single-particle energy at the Fermi surface up to a proton Fermi momentum of 
$k_F^p = 0.6$\,fm$^{-1}$, corresponding to a density of about $1.5n_0$. At higher
densities we find that the consistency begins to break down, reaching deviations of
about 10\% at $2n_0$ at which point the pairing gap vanishes.

As observed in 
Ref.\ \cite{sammarruca15} the energy per particle from the two N2LO chiral potentials
is systematically larger than that from the three N3LO potentials, and this difference grows
as the density increases. We anticipate
a corresponding increase in the $^1S_0$ proton pairing gap uncertainty band 
for densities $n \gtrsim n_0$. Beyond $n = 2 n_0$ a description of the nuclear equation 
of state based on chiral effective field theory is likely unreliable
for the low-momentum perturbative potentials considered in the present work. All results
shown below are therefore restricted to the regime $n\le 2n_0$.

\begin{figure}[t]
\includegraphics[scale=0.55]{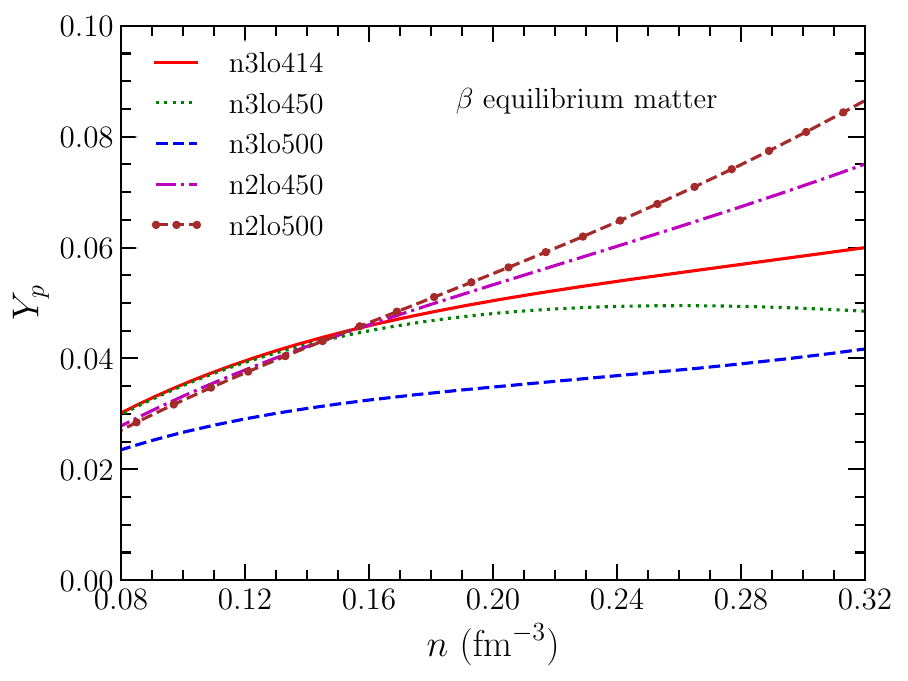}
\caption{(Color online) Proton fraction as a function of density for beta-equilibrated nuclear 
matter for $n \ge 0.5n_0$. Results are shown for the five density-dependent nuclear
interactions at N2LO and N3LO.}
\label{fig:pf}
\end{figure}

At low densities the results for the nuclear equation of state shown in Fig.\
\ref{fig:eos} are in better agreement for the different potentials. However, below 
$n \lesssim 0.5n_0$ protons in the neutron star inner crust are confined in nuclei 
and therefore do not form a macroscopic superconductor. Recently it was shown
\cite{Lim:2017luh} that the crust-core transition density $n_t$ at which unbound protons appear
lies in a limited range of $n_t \simeq 0.082-0.089$\,fm$^{-3}$ for the three N3LO chiral 
potentials considered in the present work. The transition density was identified employing
two different methods: (i) comparing the ground state energies of the homogeneous and
inhomogeneous phase as a function of density in the Thomas-Fermi approximation 
and (ii) the thermodynamic instability
method \cite{baym71} where the density of homogeneous matter is lowered until
an instability to cluster formation appears. Given the tight range of crust-core transition
densities found in Ref.\ \cite{Lim:2017luh}, we consistently take $n_t \ge 0.5n_0$
as the region above which proton pairing may occur.

\begin{figure}[t]
\includegraphics[scale=0.65]{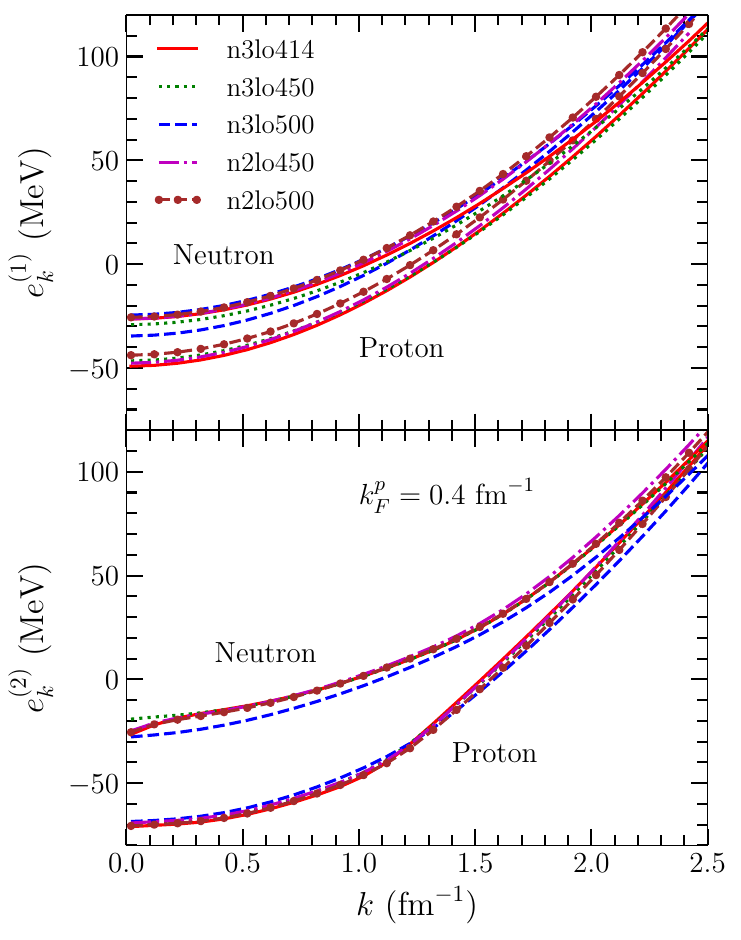}
\caption{(Color online) Single-particle energies as a function of momentum for protons 
and neutrons in beta-equilibrated nuclear matter at $k^p_F=0.4~\mathrm{fm}^{-1}$. The 
self-consistent second-order approximation to the single-particle energy, shown in Eq.\ (\ref{se2}),
is employed.}
\label{fig:spe}
\end{figure}

In Fig.\ \ref{fig:pf} we plot the proton fraction of nuclear matter in beta equilibrium as a function of
density for the five nuclear force models considered. We show only densities greater than
$n \ge 0.5n_0$ as explained above. 
Nearly all of the nuclear potentials give consistent predictions for the proton
fraction below $n < n_0$, except for the n3lo500 chiral potential which has been shown 
\cite{holt17prc} to exhibit relatively slow convergence in many-body perturbation theory.
The proton fraction in nuclear matter depends on the nuclear symmetry energy and its 
density dependence. For the n3lo500 potential the
nuclear symmetry energy is $S_v \simeq 25$\,MeV~\cite{holt17prc} when only the
first- and second-order perturbative contributions to the equation of state are included, which 
is significantly smaller than the values $S_v = 30-33$\,MeV for the other potentials considered. 
Third-order perturbative contributions have been shown \cite{holt17prc} to increase the nuclear 
symmetry energy by $2-3$\,MeV for this potential, 
but systematically including such higher-order terms in the present calculation of
the pairing gap would not meaningfully alter the final results. In all cases the values of the 
symmetry energy $S_v$ and its slope parameter $L$ are within 
the range suggested by Lattimer and Lim~\cite{Lattimer2013apj}.
Thus the proton fraction in the beta-equilibrated nuclear matter found in this work
is consistent with constraints from the most current experimental and theoretical predictions.
Beyond nuclear saturation density, the theoretical uncertainty in the proton
fraction increases significantly, and higher-order contributions
to the symmetry energy become 
important \cite{cai12,seif14,wen20}. The two N2LO chiral potentials produce the largest ground-state
energy for beta-equilibrated nuclear matter and give rise to proton fractions $Y_p = 7.5-8.5$\%
at twice saturation density. The three N3LO chiral potentials, on the other hand, predict smaller
values of $Y_p = 4-6$\% at $n = 2n_0$. 

In Fig.\ \ref{fig:spe} we show the proton and neutron single-particle energies in the Hartree-Fock
approximation $e_k^{(1)}$ (top panel) and in the self-consistent second-order
approximation $e_k^{(2)}$ (bottom panel) as a function of the momentum $k$ 
for a specific value of the proton Fermi 
momentum $k^p_F=0.4~\mathrm{fm}^{-1}$, corresponding to a total baryon
number density of $n \simeq 0.5n_0$. This is the density at which the 
neutron star inner crust transitions to homogeneous nuclear matter in the core, 
and as we show below it also corresponds to the density at which the proton $^1S_0$ 
pairing gap is maximal. We see that the inclusion of second-order perturbative corrections 
to the nucleon self energy leads to a larger isoscalar depth but also a larger isovector splitting 
between the proton and neutron single-particle energies. Quantitative inspection indicates 
that whereas the 
$e_k^{(1)}$ spectrum is nearly quadratic, and hence admits an approximation of the form in Eq.\
\eqref{eq:effm}, the $e_k^{(2)}$ spectrum deviates strongly from this form in the vicinity of the
Fermi momentum.
From Fig.\ \ref{fig:spe} we see that the different nuclear potentials give 
very similar results for the momentum dependence of the proton single-particle energy. 
As expected for the case of highly neutron-rich matter, the 
proton single-particle potential is much more strongly attractive than the neutron
single-particle potential. In fact, at the proton Fermi momentum $k_F^p = 0.4$\,fm$^{-1}$
the proton chemical potential is $\mu_p = e_p(k^p_F) \simeq -65$\,MeV. 

\begin{figure}[t]
\includegraphics[scale=0.55]{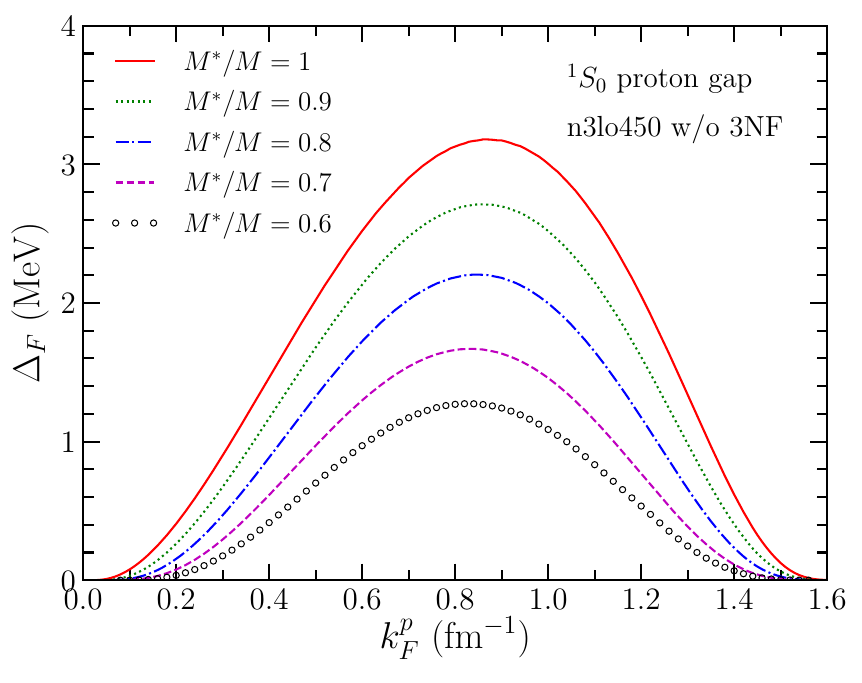}
\caption{(Color online) Density-dependent pairing gap (as a function of the proton Fermi
momentum) from chiral two-body forces. The single-particle energies in the gap equation,
Eq.\ (\ref{eq:gap}), are parametrized in terms of a density-independent effective mass
$M^*$.} 
\label{fig:n3lowo3b}
\end{figure}

We next turn our attention to the calculation of the proton pairing gap from Eq.\ (\ref{eq:gap}).
The pairing gap at the Fermi momentum $\Delta(k_F)$ is denoted by $\Delta_F$ here and
throughout.
We first neglect the presence of three-body forces, in which case the nuclear potential
is independent of the density and proton fraction, and focus on the role of the single-particle
potential, which we parametrize with different choices of the effective mass. In general,
the effective mass depends on the density and proton fraction (and also
on the momentum when the self energy is computed beyond the Hartree-Fock
approximation), but for orientation we 
consider the case of a constant effective mass. In Fig.\ \ref{fig:n3lowo3b} 
we show the proton $^1S_0$ pairing gap from the n3lo450 nucleon-nucleon potential 
as a function of the proton Fermi momentum for effective masses ranging from 
$M^*/M = 0.6-1.0$. A free proton spectrum $(M^*/M=1.0)$ gives rise to a maximum in the
pairing gap of $\Delta \simeq 3.2$\,MeV. Even a moderate reduction in the effective mass
to $M^*/M = 0.75$ leads to a decrease in the maximum of the pairing gap by a factor of 2. 
However, the density at which the pairing gap is maximal decreases by only 10\%. 
The strong dependence of the maximum in the pairing gap on the effective mass can be 
understood from Eq.\ (\ref{eq:gap}). A small effective mass corresponds to a strong 
momentum dependence of the single-particle energy around the Fermi surface. 
As the intermediate-state momentum in Eq.\ (\ref{eq:gap}) varies away from the Fermi
momentum, the energy denominator increases more rapidly for a small effective mass,
reducing the size of the pairing gap. 

\begin{table}
\caption{Proton effective masses at the Fermi surface $k_F^p=0.4$\,fm$^{-1}$ for different chiral NN + 3N interactions and two choices of the single-particle energy $e_k^{(1)}$ and $e_k^{(2)}$.}
\begin{tabular}{|c|c|c|}
\hline
$V$ & $M_p^*/M_p$\, from $e_k^{(1)}$ & $M_p^*/M_p$\, from $e_k^{(2)}$ \\ \hline
N2LO450\, & 0.76 & 0.98 \\
N2LO500\, & 0.76 & 0.97 \\
N3LO414\, & 0.80 & 0.97 \\
N3LO450\, & 0.82 & 0.93 \\
N3LO500\, & 0.80 & 0.87 \\
\hline
\end{tabular}
\label{effmtab}
\end{table}

The effective mass approximation, Eq.\ (\ref{eq:effm}), provides an accurate parametrization 
of the nucleon single-particle energy at the Hartree-Fock level. However, second-order
perturbative contributions to the nucleon self-energy lead to a strong momentum dependence 
of the effective mass that is peaked close to $M^*/M = 1$ 
at the Fermi surface \cite{bertsch68}, the regime where 
the spectrum most strongly affects the value of the pairing gap. In Table \ref{effmtab} we show the 
effective masses at the Fermi surface $k_F^p=0.4$\,fm$^{-1}$ for five different chiral NN + 3N
interactions and two choices of the single-particle energy $e_k^{(1)}$ and $e_k^{(2)}$. We see that 
the second-order perturbative corrections strongly enhance the proton effective mass in comparison 
to the Hartree-Fock values.

\begin{figure}[t]
\includegraphics[scale=0.58]{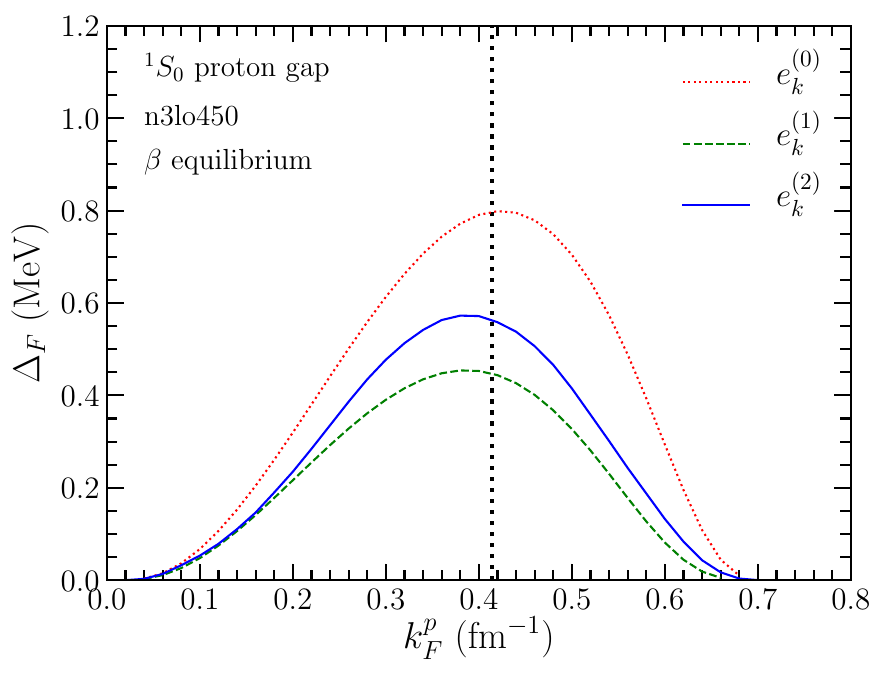}
\caption{(Color online) Proton-proton pairing gap in beta-equilibrated nuclear matter 
from the n3lo450 chiral nuclear potential, including three-body forces. 
The dotted vertical line represents the 
proton Fermi momentum at the neutron star core-crust boundary. Three approximations 
were employed for the single-particle energy spectrum: (i) free spectrum (dotted line), 
(ii) Hartree-Fock spectrum (dashed line), and (3) self-consistent second-order spectrum
(solid line).}
\label{fig:effmspe}
\end{figure}

In Fig.\ \ref{fig:effmspe} we 
study the effect of different parametrizations of the nucleon single-particle energy on the 
density-dependent pairing gap. In all cases we include both two- and three-body forces. 
In the first case, shown as the dotted curve in Fig.\ \ref{fig:effmspe}, we consider a 
free-particle spectrum $e_k^{(0)} = k^2/2M$. 
The dotted vertical line stands for
the Fermi momentum at the core-crust boundary of a neutron 
star ($n \sim 1/2n_0,\, Y_p\sim 0.03$). Comparing to Fig.\ \ref{fig:n3lowo3b} we see 
that three-body forces lead to a reduction in the maximum value of the pairing gap by a 
factor of four. Although the proton Fermi momentum is small, the large 
neutron density leads to a more strongly repulsive effective two-body proton-proton interaction
as shown in Ref.\ \cite{Holt10}. Consequently the maximum proton pairing gap shown 
in Fig.\ \ref{fig:effmspe} is roughly $1/3$ the $^1S_0$ neutron pairing gap in neutron star inner 
crusts \cite{Maurizio14}, where three-body forces play a much smaller role. Treating
the single-particle energy in the Hartree-Fock approximation $e_k^{(1)} = k^2/2M + 
\Sigma^{(1)}(k)$ leads to an additional reduction in the pairing gap by about 40\% as 
shown by the dashed line of Fig.\ \ref{fig:effmspe}. Finally, employing the self-consistent
second-order single-particle energy $e^{(2)}_k$ (see Eq.\ (\ref{se2})) in the denominator
of the gap equation leads to an increase of 20\% in the maximum gap size relative to
the Hartree-Fock approximation. This may be understood from the fact 
that the second-order contribution $\Sigma^{(2)}(e_k,k)$ to the self energy on average 
increases the effective mass in the vicinity of the Fermi surface.

\begin{figure}[t]
\includegraphics[scale=0.58]{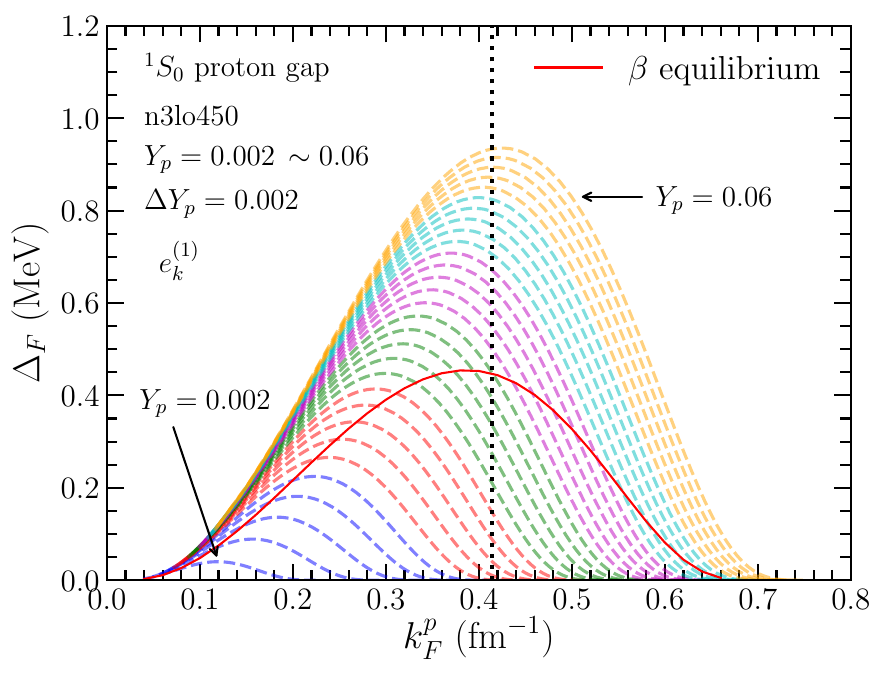}
\caption{(Color online) Density-dependent proton-proton pairing gap 
from the n3lo450 chiral nuclear potential 
for different values of the proton fraction $Y_p$ and for nuclear matter in beta equilibrium. A
Hartree-Fock single-particle spectrum is employed.}
\label{fig:n3lowi3beff}
\end{figure}

Fig.~\ref{fig:n3lowi3beff} shows the $^{1}S_0$ proton pairing gap in the presence of 
three body forces using the n3lo450 chiral nuclear potential.
The dashed curves correspond to different values of the (fixed) proton fraction $Y_p$, which
ranges from $0.002 \le Y_p \le 0.06$ with $\Delta Y_p = 0.002$, 
and the solid red curve is that for nuclear matter in beta equilibrium. 
For a given $Y_p$ we calculate the solution to the BCS gap equation using the first-order
approximation for the single-particle energies $e^{(1)}(k)$. We see that the proton fraction is
an important parameter for determining the size of the pairing gap. 
For instance at $k_F^p = 0.4$\,fm$^{-1}$, changing the proton fraction from $Y_p = 0.03$ to
$Y_p = 0.04$ would increase the gap size from $\Delta_F = 0.5$\,MeV to $\Delta_F = 0.75$\,MeV.

We note that the nuclear potential $V_{\rm eff}(k,k^\prime)$ depends on the proton fraction
when three-body forces are included. As shown in Fig.~\ref{fig:n3lowi3beff},
the proton pairing gap and the available pairing domain in $k_F^p$ increase as
the proton fraction increases because $V_{\rm eff}(k,k^\prime)$ depends sensitively on
the proton fraction. As mentioned in Section \ref{sec:int}, three-body forces have
been considered previously in a phenomenological way to compute the proton pairing 
gap in beta-stable nuclear matter.
In this work, three-nucleon forces consistent with the low-energy constants in the two-body
force and fitted to the properties of $A=3$ nuclei have been employed.
In addition we have calculated the nuclear EOS with the same nuclear forces to determine
the proton fraction.

\begin{figure}[t]
\includegraphics[scale=0.55]{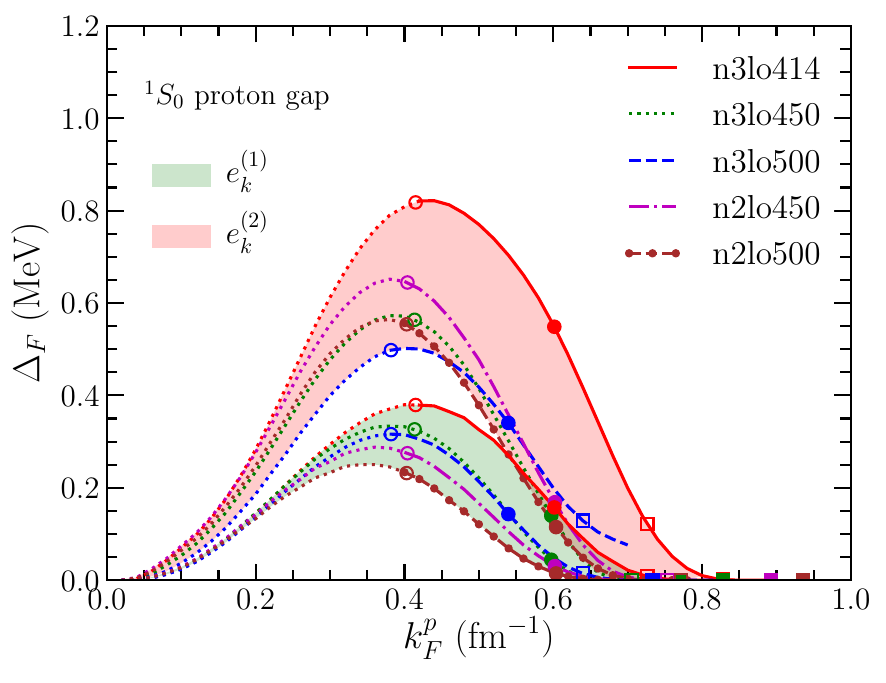}
\caption{Proton $^1S_0$ pairing gap as a function of the Fermi momentum $k_F$ for the
five chiral potentials considered in the present work.}
\label{fig:gapspe}
\end{figure}

Fig.~\ref{fig:gapspe} shows the proton pairing gap in beta-equilibrium matter
 using five different chiral potentials and two
different approximations for the single-particle spectrum: $e_k^{(1)}$ (green) and $e_k^{(2)}$ (red).
The dotted sections of the curves indicate the pairing gap for densities lower than that
of the neutron star core-crust boundary. The large symbols on the curves indicate the values 
of the pairing gap at nuclear densities $n=n_0/2$ (open circle), $n_0$ (filled circle), 
$3n_0/2$ (open square), and $2n_0$ (filled square). Only the proton pairing gap from
the N3LO500 potential using the $e_k^{(2)}$ spectrum does not show closure in $k_F^p$. 
This is due to the associated small proton fraction (see Fig.\ \ref{fig:pf}), which leads to a larger
value of the total baryon number density for a given value of $k_F^p$. We have restricted
our calculations to the density regime $n \le 2n_0$ and therefore do not report results 
for $k_F^p > 0.72$\,fm$^{-1}$ from the N3LO500 potential.
In particular, just below twice saturation density, the self-consistent calculation of 
the single-particle potential from the N3LO500 potential begins to break down.

We note several nearly universal features in the results of Fig.~\ref{fig:gapspe}, 
independent of the choice of chiral interaction and the associated derived
quantities, such as the single-particle spectrum and proton fraction. First, for all cases the peak in 
the pairing gap occurs very close to the crust-core boundary and within a very small window of 
the proton Fermi momentum $0.35\,{\rm fm}^{-1} < k_F^p < 0.43\,{\rm fm}^{-1}$. Second, apart 
from the results of the N3LO414 and N3LO500 potentials using the $e_k^{(2)}$ energy spectrum, 
nearly all potentials lead to a proton pairing gap that vanishes when the proton Fermi momentum 
is in the range $0.65\,{\rm fm}^{-1} < k_F^p < 0.75\,{\rm fm}^{-1}$. Even the inclusion of both the N3LO414 and N3LO500 potentials would only increase the upper bound to 
$k_F^p \simeq 0.8\,{\rm fm}^{-1}$, which corresponds to a total baryon number density less than
$2n_0$.  We therefore note that proton pairing is 
expected to exist within a neutron star at densities where chiral effective field theory 
is valid. 

Chiral potentials generally become more repulsive as the momentum-space
cutoff $\Lambda$ increases, which partly accounts for the smaller proton pairing gaps associated with
the N2LO500 and N3LO500 chiral potentials and the largest proton pairing gaps associated with
the N3LO414 potential.
For neutron matter and beta stable nuclear matter, 
it was also shown that the N2LO equations of state are stiffer than at N3LO. The effect of
repulsive contributions in the nuclear potential are mostly clearly seen in the pairing gaps 
associated with the $e_k^{(1)}$ spectrum, where there is a clear trend from the N2LO
potentials (with the smallest gaps) to the N3LO potentials ordered according to the value of
the cutoff $\Lambda$. In addition to the EOS stiffness (where an 
attractive force would increase the gap size and the EOS would be soft), the gap size is
also related to the proton 
fraction in beta-equilibrium matter (which controls the proton Fermi momentum),
and the proton single-particle spectrum.
From Fig.~\ref{fig:gapspe} we see that the inclusion of second-order contributions 
to the single-particle energy has a strong impact on the proton pairing gap. 
The nucleon effective mass at the Hartree-Fock $e_k^{(1)}$
approximation is small $M^*/M \sim 0.75$, while second-order perturbative contributions 
lead to a strong
energy dependence in the single-particle potential that increases the effective mass to 
$M^*/M \sim 1$ near the Fermi surface. As can be inferred from Fig.\ \ref{fig:n3lowo3b}, this 
generically leads to larger proton pairing gaps in Fig.\ \ref{fig:gapspe} for the $e_k^{(2)}$ spectrum.
Specifically, in the $e_k^{(2)}$ approximation we find that the N2LO450, N2LO500, and 
N3LO414 potentials experience the largest changes in the proton effective mass, which 
enhances the magnitude of the pairing gap relative to their values with the $e_k^{(1)}$ spectrum.
From the above considerations we find that multiple effects strengthen
the pairing gap associated with the N3LO414 potential, which explains why it deviates most strongly
from the other potentials when the $e_k^{(2)}$ spectrum is employed.

\begin{figure}[t]
\includegraphics[scale=0.55]{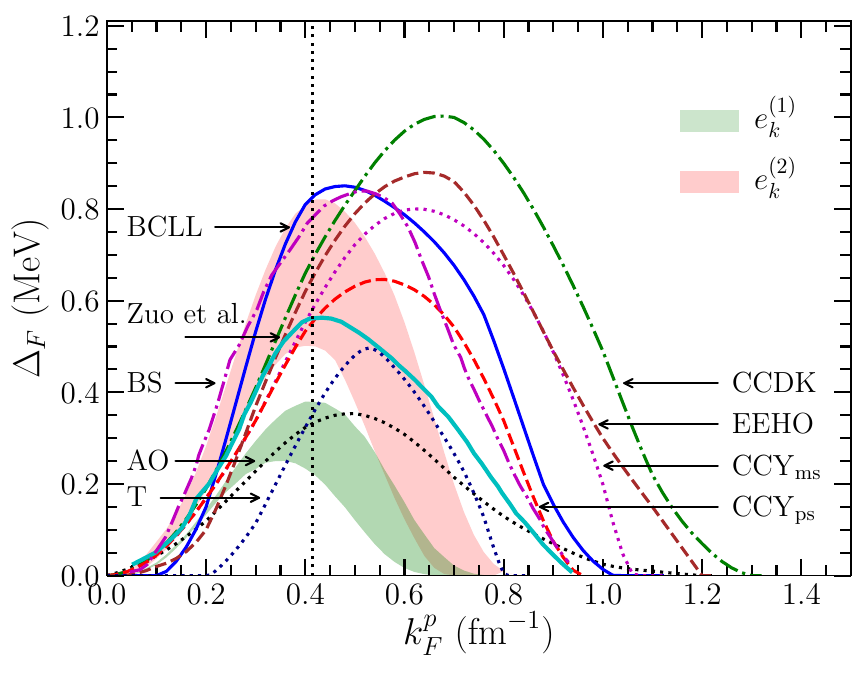}
\caption{(Color online) $^{1}S_0$ proton pairing curves from nuclear model. 
Our EFT calculation (red band) gives similar pairing gaps and smaller range of Fermi momentum where
the pairing is available.
For comparison, it is shown the previous BCS calculations, 
Chao et al.~\cite{CHAO1972320}(`CCY'),
Takatsuka~\cite{Tak73}~(`T'),
Amundsen and {\O}stgaard~\cite{Amundsen:1984qq}~(`AO'),
Baldo et al.~\cite{BALDO1992349}~(`BCLL'),
Chen et al.~\cite{CHEN199359}~(`CCDK'),
Elgar{\o}y et al.~\cite{Elgaroy:1996mx}~(`EEHO'),
Zuo et al.~\cite{ZUO200444}(`Zuo et al.'),
and
Baldo and Schulze~\cite{Baldo07}(`BS').
}
\label{fig:comp_all}
\end{figure}

In Fig.\ \ref{fig:comp_all} we compare the proton pairing gap uncertainty band calculated
in the present work to previous results in the literature.
We find that employing the $e_k^{(2)}$ approximation
for the single-particle spectrum, the maximum in the pairing gap lies
in the range $0.51\,{\rm MeV} < \Delta_F < 0.82$\,MeV, which is consistent with previous
calculations, but the maximum density at which proton pairing is expected to occur is
systematically smaller than other models. This is largely caused by three-body forces and 
the behavior of the chiral potential $V_{\rm eff}(k,k^\prime)$ as the proton fraction is
increased in neutron star matter. Employing the $e_k^{(1)}$ spectrum we find instead
that $0.25\,{\rm MeV} < \Delta_F < 0.38$\,MeV. The medium-dependent nuclear potential in 
isospin-asymmetric nuclear matter might also affect to a lesser extent the 
${}^3P_2-{}^3F_2$ neutron pairing gap, which is typically 
calculated in pure neutron matter. We note that our error bands in Fig.\ \ref{fig:comp_all}
partially account for uncertainties due to (i) the convergence of the chiral expansion 
(where we have varied the chiral order of the nucleon-nucleon interaction from N2LO to N3LO),
(ii) the choice of the resolution scale at which nuclear dynamics is resolved (encoded in
the high-momentum cutoff in the chiral potential), and (iii) the convergence in the many-body 
expansion (through different choices of the single-particle spectrum). A more comprehensive
account of uncertainties would include varying the chiral EFT low-energy constants within
ranges consistent with 2N and 3N scattering data and three-body bound state properties
\cite{Wesolowski:2018lzj,Reinert:2017usi}, 
improved order-by-order effective field theory truncation errors 
\cite{Epelbaum:2014efa,Furnstahl:2015rha,Drischler:2020hwi} including consistent N3LO
three-body forces 
\cite{ishikawa07,bernard08,bernard11,tews13,Drischler:2016djf,Holt:2019bah}, 
and the improved description of short- and long-range correlations 
that go beyond the BCS approximation \cite{Page:2013hxa,Ding16}.

In the weak coupling approximation, the critical temperature for the onset of pairing is
given by \cite{LPSTAT}
\begin{equation}
T_c \simeq 0.57\, \Delta_F(T=0).
\end{equation} 
We find that in the present analysis with the $e_k^{(2)}$ spectrum, the critical
temperature for proton pairing in the core of neutron stars is 
\begin{equation}
T_c \sim (3.2 - 5.1)\times 10^{9} \,\mathrm{K}\,.
\end{equation}
Compared to the range of critical temperatures
predicted in a recent study from neutron star cooling
using Bayesian analysis~\cite{Beloin:2016zop},
where $T_c = 7.59_{-5.81}^{+2.48} \times 10^9~\text{K}$, our prediction
has a smaller central value but is consistent at the highest range.

\begin{figure}[t]
	\centering
	\includegraphics[scale=0.55]{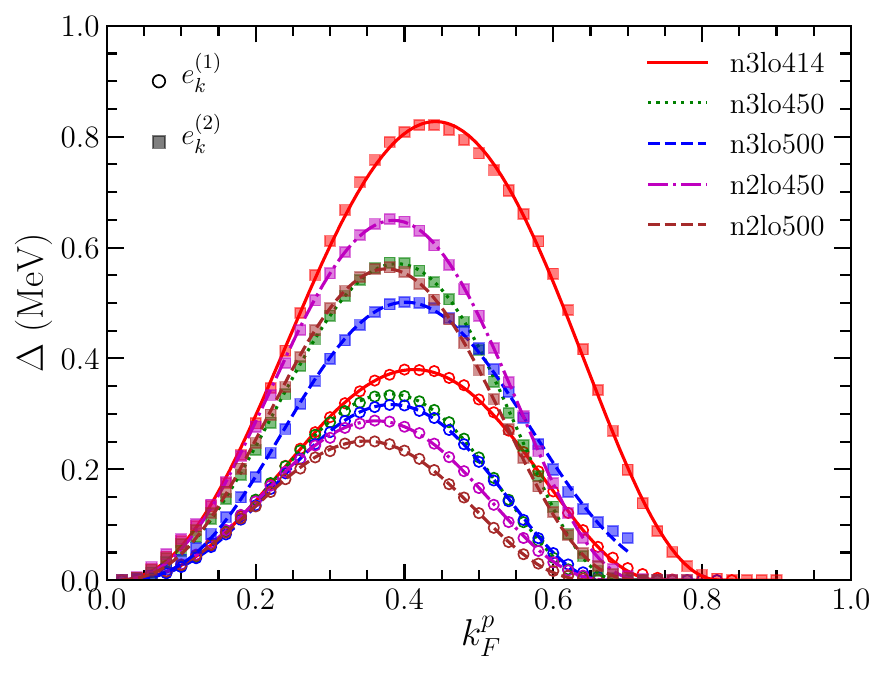}
	\caption{Calculated proton pairing gaps using the $e_k^{(1)}$ single-particle
	energy (open symbols) and the $e_k^{(2)}$ single-particle energy (filled symbols) together
	with the parameterization (solid lines) in Eq.\ \eqref{param}.}
	\label{fig:eftspehf2ndgapfitcomp}
\end{figure}

Finally, we consider parameterized fitting functions for the pairing gaps shown in 
Fig.\ \ref{fig:gapspe}. We find that the pairing gap can be well fitted with the 
simple function\,\cite{Lim:2015lia}
\begin{equation}
\Delta(k_F) = 
\begin{cases}
\Delta_{\rm m} \mathcal{N}(k_F-k_1)^\alpha (k_2-k_F)^\beta,
&\text{if} \,\,k_1 < k_f <k_2\,, \\
0, & \text{otherwise}\,,
\end{cases}
\label{param}
\end{equation}
where $\mathcal{N}$ is the normalization factor given by
\begin{equation}
\mathcal{N} = \frac{1}{\alpha^\alpha \beta^\beta}
\left( \frac{\alpha +\beta}{k_2-k_1}\right)^{\alpha+\beta}\,.
\end{equation}
In Fig.\ \ref{fig:eftspehf2ndgapfitcomp} we plot the numerical pairing gaps as well
as the fitting functions under the two approximations $e_k^{(1)}$ (open circles)
and $e_k^{(2)}$ (filled squares) for the single-particle energies. We see that in all
cases the parameterized form in Eq.\ \eqref{param} captures very well the $k_F^p$
dependence of the gaps. In Table \ref{paramt} will list the values of the parameters
for the different chiral interactions and choices of single-particle spectrum.

\begin{table}[t]
	\caption{Fitting parameters for the $^{1}S_0$ proton pairing gap in Eq.\ \eqref{param} for
	different chiral potentials and choices for the single-particle energy spectrum. }
	\begin{tabular}{cccccc}
		\hline
		\hline

\multirow{2}{*}
        {Model} & $\Delta_{\rm m}$ & $k_1\,$                      & $k_2\ $                       & $\alpha$ &$\beta$\\
		            & (MeV)                     & $(\mathrm{fm}^{-1})$ & $(\mathrm{fm}^{-1})$ & & \\
\hline
$e_k^{(1)}$ n3lo414 & 0.380 &  0.000   &    0.748     & 2.998      &   2.432   \\
$e_k^{(1)}$ n3lo450 & 0.336 &  0.000   &    0.663      & 3.116     &  2.319       \\
$e_k^{(1)}$ n3lo500 & 0.317 &  0.004   &    0.672      & 3.041     &  2.325       \\
$e_k^{(1)}$ n2lo450 & 0.287  &  0.000   &    0.682       & 3.058    &   2.713       \\
$e_k^{(1)}$ n2lo500 & 0.251  &  0.000   &    0.658      & 3.054      & 2.723       \\
\hline
$e_k^{(2)}$ n3lo414 & 0.827 & 0.002   &    0.828      &  2.769      &  2.434   \\
$e_k^{(2)}$ n3lo450 & 0.571 & 0.000   &    0.689      &  2.965      &  2.275   \\
$e_k^{(2)}$ n3lo500 & 0.501 & 0.000  &     0.860      &  3.507      &   4.003  \\
$e_k^{(2)}$ n2lo450 & 0.649 & 0.000  &     0.726      &  3.019      &  2.672   \\
$e_k^{(2)}$ n2lo500 & 0.562 & 0.000  &     0.722      &  3.159      & 2.914    \\
\hline
		\hline
	\end{tabular}
	\label{paramt}
\end{table}

\section{Summary}
In this work we have studied the proton $^{1}S_0$ pairing gap in nuclear matter at
beta equilibrium using five different 
nuclear two- and three-body potentials derived within the framework of chiral effective
field theory. Nucleon-nucleon potentials at both N2LO and N3LO were considered, together
with the chiral three-body force at N2LO. In addition to the choice of nuclear potential,
also the single-particle spectrum employed in the BCS gap equation is a source of
theoretical uncertainty.

We find that both three-body forces and a realistic proton single-particle
potential in neutron star matter reduce the maximum size of the proton $^1S_0$
pairing gap. In particular, three-body forces reduce the maximum gap size by a factor
of 3, while a self-consistent second-order treatment of the proton single-particle potential 
leads to an additional reduction of about 30\%. Our results for the  
$^{1}S_0$ proton pairing gap have a similar range of sizes compared to previous 
studies. However, the maximum density at which proton pairing may exist in neutron
stars is systematically smaller than previous results. This ultimately comes from 
the inclusion of three-body forces in our effective field theory calculation, which requires
a consistent calculation of the proton fraction in beta-equilibrium matter. 
The three-body force leads to additional repulsion in the effective interaction and
a suppression in the pairing gap as the density increases.

These results will be important for a consistent description of neutron star cooling.
Proton ${}^{1}S_0$ pairing will likely not give any reduction factor for nucleon direct 
Urca cooling, since the pairing gap is seen to vanish well before the proton fraction 
reaches a value high enough for the onset of the direct URCA process.
However, proton pairing will certainly give a reduction factor to the thermal conductivity,
heat capacity, and neutrino emission processes involving protons. 
Thus the enhanced cooling processes in neutrons stars arising from Cooper-pair
breaking/formation is likely to be dominated by $^{3}P_2$ neutron pairing in the core
rather than $^1S_0$ pairing of protons.

\acknowledgments

Work supported by the National Science Foundation under Grant No.\ PHY1652199 and 
by the U.\ S.\ Department of Energy National
Nuclear Security Administration under Grant No.\ DE-NA0003841.
Y.\ Lim has received support by the Max Planck Society and the Deutsche
Forschungsgemeinschaft (DFG, German Research Foundation) -- Project ID 279384907 -- SFB 
1245. Portions of this research were conducted with the advanced computing resources provided 
by Texas A\&M 
High Performance Research Computing.

%

\end{document}